\documentclass[aps,prb,notitlepage,twocolumn,superscriptaddress]{revtex4-2}

\makeatletter
\def\@hangfrom@section#1#2#3{\normalsize\@hangfrom{#1#2}#3}
\def\@hangfroms@section#1#2{\normalsize#1#2}
\makeatother

\RequirePackage[usenames, dvipsnames, x11names]{xcolor}
\usepackage{amsmath,amsfonts,amssymb,bm}
\usepackage{graphicx,color,soul}
\usepackage{dsfont}
\usepackage{enumerate}
\usepackage{multirow}

\usepackage[none]{hyphenat} 
\usepackage{tabularx}

\usepackage{newtxtext,newtxmath}
\usepackage{hyperref}
\hypersetup{colorlinks=true,
                    linkcolor=blue,
                    urlcolor=blue,
                    citecolor=blue}

\renewcommand{\addcontentsline}[3]{}

\begin{document}
\graphicspath{{figure/}}

\title{Site-selective renormalization and competing magnetic instabilities in paramagnet Y$_3$Cu$_2$Sb$_3$O$_{14}$}

\author{Yanpeng Zhou}
\affiliation{\mbox{State Key Laboratory of Quantum Functional Materials, SPST, ShanghaiTech University, Shanghai 201210, China}}
\author{Gang Li}
\email{ligang@shanghaitech.edu.cn}
\affiliation{\mbox{State Key Laboratory of Quantum Functional Materials, SPST, ShanghaiTech University, Shanghai 201210, China}}
\affiliation{\mbox{ShanghaiTech Laboratory for Topological Physics, ShanghaiTech University, Shanghai 201210, China}}

\date{\today}
\begin{abstract}
Quantum spin liquids (QSLs) are exotic phases of matter characterized by long-range entanglement and the absence of magnetic order even at zero temperature.
Here, we present a comprehensive theoretical study of the frustrated magnet Y$_3$Cu$_2$Sb$_3$O$_{14}$ to elucidate its electronic and magnetic properties. 
We uncover completely opposite crystal-field splittings of the two inequivalent Cu sites owing to their fundamentally distinct oxygen coordination—trigonal distorted octahedral CuO$_6$ and axially compressed CuO$_8$.
This inversion places the unpaired hole in the $d_{z^2}$ orbital at the Cu-2 site, while Cu-1 maintains conventional $d_{x^2-y^2}/d_{xy}$ character, which results in a selective band-renormalization of orbitals from the two Cu ions.  
We further find multiple magnetic instabilities competing with nearly equal strength in this system: the spin susceptibility lacks dominant peaks, and the leading eigenvalues approach unity simultaneously across all wavevectors with increasing interactions. 
This competitive interplay, originating from the distinct local environments and geometric frustration on the triangular lattice, agrees well with the absence of long-range magnetic order in experiment. 
Our results support Y$_3$Cu$_2$Sb$_3$O$_{14}$ as a promising QSL candidate where the unique combination of disparate crystal-field environments, strong correlations, and competing exchange interactions conspire to stabilize an exotic quantum ground state.
\end{abstract}
\maketitle

\section{\label{sec:1}Introduction}
The pursuit of the quantum spin liquid (QSL) state—a highly entangled phase where magnetic moments remain dynamic at absolute zero—is a cornerstone of modern condensed matter physics \cite{anderson1973}. Theoretically, the antiferromagnetic triangular lattice is the canonical platform for such a phase, where geometrical frustration precludes simple N\'{e}el order. While the pure $J_1$ isotropic Heisenberg model on a triangular lattice eventually settles into a 120$^\circ$ ordered state \cite{capriotti1999}, the introduction of competing next-nearest-neighbor interactions ($J_2/J_1 \sim ( 0.06, 0.17))$ is predicted to stabilize exotic gapped or gapless QSL phases \cite{zhu2015, PhysRevB.92.140403, PhysRevB.93.144411}.

Despite decades of experimental searching, a ``firmly accepted'' triangular QSL remains elusive due to persistent material-specific challenges. First, structural disorder, such as the Mg/Ga site-mixing in the widely studied YbMgGaO$_4$, can mimic QSL signatures by inducing a random singlet or spin-glass state rather than a true QSL phase~\cite{paddison2017, PhysRevLett.118.107202, PhysRevLett.117.267202, PhysRevLett.120.087201, xfsun-2021}. 
Second, many 2D candidates suffer from subtle interlayer couplings or exchange anisotropies that drive the system toward long-range order at sub-Kelvin temperatures, as seen in various delafossite-related structures~\cite{liu2018}.
Among them,  alkali-based $A$Yb$Ch_2$ ($A$ = Na, K; $Ch$ = O, S, Se) family has been extensively studied. 
Most notably, NaYbSe$_2$ has emerged as a premier QSL candidate due to its lack of long-range magnetic ordering and the presence of coherent, fractionalized spinon excitations. 
Early studies proposed a gapless spinon Fermi surface ground state~\cite{PhysRevB.100.144432, PhysRevX.11.021044, PhysRevB.110.224414} from ultralow temperature susceptibility, specific heat, and thermal conductivity measurements. 
Similarly, TlYbSe$_2$ has emerged as a rare example of a field-tunable QSL where gauge field excitations may play a critical role~\cite{2025arXiv250405436B}.
Furthermore, the discovery of PrMgAl$_{11}$O$_{19}$ has introduced the possibility of an Ising-type QSL, expanding the search beyond isotropic Heisenberg symmetries~\cite{ASHTAR2019146, C9TC02643F}.
While the ``Ising-type" correlation and the absence of magnetic order are firmly established across all studies~\cite{PhysRevB.110.134401, Cao_2024, PhysRevB.109.165143, PhysRevResearch.6.043147}, there is currently an active debate regarding whether the QSL ground state is gapless or gapped~\cite{PhysRevB.109.165143, PhysRevResearch.6.033031,PhysRevResearch.6.043147}. 
Among the triangular systems, there is also a class of materials named cluster Mott insulators, where molecular units rather than individual atoms carry the effective spin-1/2 moments. 
The Mo$_{3}$O$_{8}$-based kagome systems (LiZn$_{2}$Mo$_{3}$O$_{8}$ family)~\cite{cotton_metal_1964, mccarroll_structural_1977, cotton_preparations_1991, sheckelton_possible_2012, flint_emergent_2013, mourigal_molecular_2014, chen_cluster_2016, chen_emergent_2018}, and Nb$_3$X$_8$ family (X = Cl, Br, I)~\cite{khvorykh_niobium_1995,kennedy_chemical_1991, kennedy_experimental_1996,sheckelton_rearrangement_2017, haraguchi_magneticnonmagnetic_2017, miller_solid_1995, pasco_tunable_2019, nanolett_feng_2022, hu2023correlated, grytsiuk_nb3cl8_2024, gao_discovery_2023,  aretz_strong_2025, liu_direct_2025} exemplify this class, where trimerization leads to effective triangular lattices of spin clusters. 
The $1T$-TaS$_{2}$ system similarly features star-of-David clusters forming a triangular arrangement~~\cite{TaSe2_cluster, 2010Natur.464..199B, TaS2-nat.m, TaSe2-nat.p,tas2-Law2017,tas2-He2018,tas2-Shi2021,tas2-Shen2022,tas2-Ruan2021,tas2-Tian2024,tas2-Wang2020,tas2-Manas-Valero2021}. 
These molecular QSL candidates bridge the gap between atomic-scale frustrated magnets and emergent lattice geometries.

Despite these advances, characterizing the true ground state remains difficult. 
The lack of a clear ``smoking gun" signature often lead to ambiguous results. 
In this sense, new candidate materials may provide valuable information from different perspective and fundamentally improve the understanding of this exotic state.  
In this context, Y$_3$Cu$_2$Sb$_3$O$_{14}$ offers a unique advantage~\cite{saha2025}. 
Featuring a 3D network of two inequivalent Cu$^{2+}$ ($S$=1/2) triangular layers, it possesses an architectural robustness that avoids the site-mixing issues of mixed-cation systems. 
Unlike the 2D candidates mentioned above, Y$_3$Cu$_2$Sb$_3$O$_{14}$ exhibits a distinct two-step spin condensation—first at 120~K and again below 1~K—revealing a complex hierarchy of fractionalized excitations~\cite{saha2025}. Magnetic susceptibility and muon spin relaxation ($\mu$SR) confirm that the system remains dynamic down to 0.077~K,  significantly below the interaction scale ($\Theta_{CW} \approx -20$~K), which positions it as a superior platform for exploring the intrinsic physics of three-dimensional quantum frustration. 
While experimental study in Ref.~\cite{saha2025} has revealed several promising features of Y$_3$Cu$_2$Sb$_3$O$_{14}$, a theoretical understanding of the microscopic electronic and magnetic dynamics of this system is still missing. 

In this work, we present a theoretical investigation of the spin-$1/2$ three-dimensional frustrated magnet Y$_3$Cu$_2$Sb$_3$O$_{14}$. By combining density-functional theory (DFT) calculations, perturbative many-body methods, and dynamical mean-field theory (DMFT), we analyze the electronic and magnetic responses of this material, offering insight into its unusual low-energy excitations. 
The non-magnetic electronic structure is governed by two inequivalent Cu octahedra that exhibit opposite trigonal distortion modes, leading to fundamentally different crystal-field splittings. 
The low-energy degrees of freedom are described by narrow bands, which are renormalized by electronic correlations to different extent featuring a unique site-selective character. 
Calculations of the spin susceptibility reveal a strong competition among multiple magnetic wave vectors, suggesting that the ground state may realize a quantum spin liquid.

The remainder of this paper is organized as follows. In Section~\ref{sec:results-and-analysis}, we first discuss the electronic structure obtained from DFT and DFT+DMFT, demonstrating the unique crystal-field of the two Cu ions and electronic band renormalization. 
We then examine the spin susceptibility within the fluctuation-exchange approximation (FLEX), confirming its competitive nature and the absence of a dominant magnetic instability. 
Finally, in Section~\ref{sec:discussions-and-conclusions}, we summarize our conclusions and discuss their implications for experimental identification of a quantum spin liquid state in Y$_3$Cu$_2$Sb$_3$O$_{14}$.

\section{Results}\label{sec:results-and-analysis}

\subsection{Inverted crystal-field splittings}

\begin{figure*}[t]
	\centering
	\includegraphics[width=\textwidth]{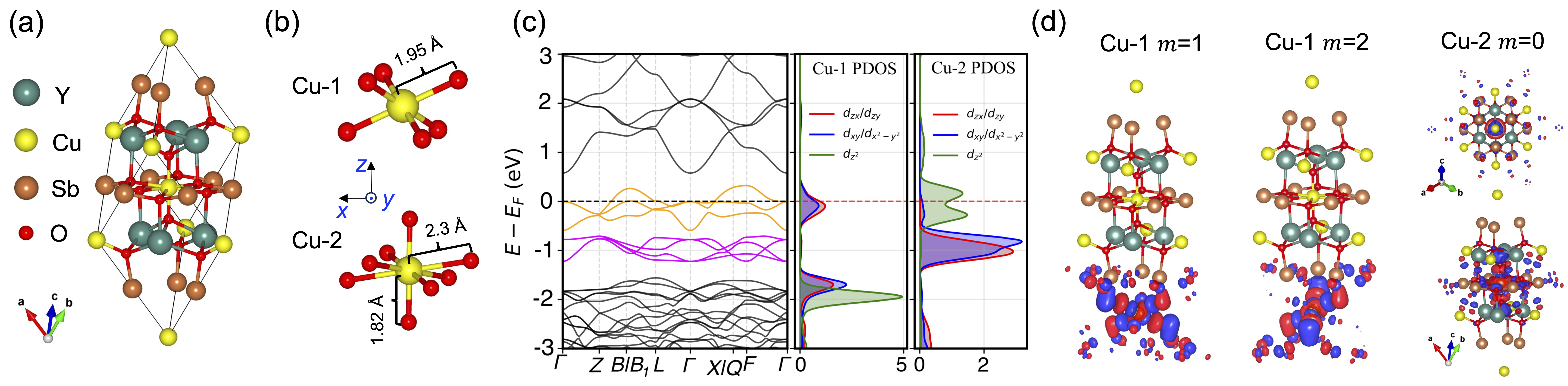}
	\caption{\textbf{Crystal and electronic structure of Y$_3$Cu$_2$Sb$_3$O$_{14}$.} (a) The crystal structure of Y$_3$Cu$_2$Sb$_3$O$_{14}$, in which the two inequivalent copper sites (yellow atoms), Cu-1 and Cu-2, individually form triangular planes. (b) shows the two different local environment of Cu-1 and Cu-2. In Cu-2, an additional  shorter Cu-O bond along z axis inverts the crystal field created by the regular trigonal distortion in Cu-1. 
	(c) The DFT electronic band structure of the primitive cell is shown along a high-symmetry path in the first BZ. The orange line around the Fermi level denotes the three bands with their Wannier functions shown in (d). Additionally including the purple bands leads to a ten bands effective model. Both the three-band and the ten-band models show excellent agreement with the DFT results. (d) The projected Wannier functions of the three yellow bands indicate that, among the three orbitals,  two orbitals are from the Cu-1 ion and one from Cu-2 ion.}
	\label{fig:dft-result}
\end{figure*}

Y$_3$Cu$_2$Sb$_3$O$_{14}$ crystallizes in the $R\bar{3}m$ structure~\cite{saha2025}, illustrated in Fig.~\ref{fig:dft-result}(a). 
 The structure comprises interconnected polyhedra of copper and antimony coordinated by oxygen, with yttrium cations occupying interstitial sites to maintain charge balance. 
 The three-dimensional framework features two inequivalent Cu$^{2+}$ ions, each forming edge-shared triangular lattices. This geometry gives rise to strong magnetic frustration and deviates from conventional ordered magnetic arrangements.

The two inequivalent Cu sites experience markedly different oxygen coordination environments: CuO$_6$ and CuO$_8$. Each Cu sublattice forms an in-plane triangular lattice with a layered geometry. For clarity, we denote the Cu atom at the corner of the unit cell $(0,0,0)$ as Cu-1 and the one at the center $(1/2,1/2,1/2)$ as Cu-2. At the Cu-1 site, each Cu atom is coordinated by six oxygen atoms with equal bond lengths of 1.95~\AA, but the bond angles deviate significantly from $90^\circ$, resulting in a compressive trigonal distortion. This coordination departs from a regular octahedral environment and instead exhibits local symmetry close to $D_{3d}$. Under such a crystal field, the fivefold-degenerate $d$ orbitals split into three distinct manifolds: a nondegenerate $A_{1g}$ orbital (predominantly $d_{z^2}$) and two sets of doubly degenerate $E_g$ levels, with the typical energy ordering
\[
E(d_{z^{2}})<E(d_{x^{2}-y^{2}},d_{xy})<E(d_{xz},d_{yz}).
\]

In contrast, Cu-2 is surrounded by eight oxygen atoms, forming a distinct local environment. This site is characterized by two different Cu–O bond lengths: a short axial distance of 1.82~\AA ~along the $C_3$ axis and longer equatorial distances of 2.3~\AA ~within the two triangular planes. The strong axial compression along the threefold axis generates an unconventional crystal-field splitting. The short axial bonds produce pronounced $\sigma$-antibonding interactions that selectively destabilize the $d_{z^2}$ orbital (transforming as $A_{1g}$), pushing it above the other $d$ states. The resulting energy ordering is
\[
E(d_{xz},d_{yz})<E(d_{x^{2}-y^{2}},d_{xy})<E(d_{z^{2}}),
\]
with orbitals transforming as $E_g$, $E_g$, and $A_{1g}$ under $D_{3d}$, respectively. This inversion of the conventional level scheme places the unpaired hole of Cu$^{2+}$ in the $d_{z^2}$ orbital, establishing a half-filled local configuration. 
The inverted crystal-field splitting and the partial occupancy of the Cu-1 and Cu-2 $d$ orbitals predict a metallic ground state of this material, which is confirmed by the DFT electronic structure calculations discussed below. 

\subsection{DFT electronic structure}

Having established the local valence configuration and crystal field, we now examine the Bloch band structure of Y$_3$Cu$_2$Sb$_3$O$_{14}$. 
Figure~\ref{fig:dft-result}(c) shows the electronic dispersion along a chosen high-symmetry path in the Brillouin zone (BZ).
DFT predicts a metallic ground state, with Cu-$d$ orbitals dominating the low-energy states near the Fermi level. The projected density of states (PDOS) in the right panel of Fig.~\ref{fig:dft-result}(c) reveals distinct $d$-orbital contributions from Cu-1 and Cu-2. Consistent with the crystal-field analysis above, the two sites share the same local $D_{3d}$ point-group symmetry, but the shorter axial Cu–O bonds at Cu-2 induce a complete inversion of the $d$-orbital energy ordering relative to Cu-1. At Cu-1, the $d_{z^2}$ orbital lies at approximately $-2$ eV as the lowest among the five $d$ states, whereas at Cu-2 it becomes the highest. Consequently, Cu-1 contributes a twofold $E_g$ manifold, while Cu-2 provides a nondegenerate $A_{1g}$ state to the Fermi surface, suggesting that the minimal low-energy model must include at least three orbitals.

Using maximally localized Wannier functions as implemented in Wannier90~\cite{pizzi2020wannier90}, we construct a three-orbital model (yellow solid line in Fig.~\ref{fig:dft-result}(c)) that reasonably reproduces the DFT bands. 
The corresponding Wannier functions, displayed in Fig.~\ref{fig:dft-result}(d), exhibit two distinct symmetry types. Two of the Wannier functions reside on Cu-1 and reflect the twofold $E_g$ symmetry, while the third is centered on Cu-2 and displays $C_3$-symmetric $A_{1g}$ character, fully consistent with the two distinct $D_{3d}$ crystal-field splittings.
Three $d$ valence electrons occupy the two $E_g$ Wannier functions at Cu-1 while the $A_{1g}$ Wannier function at Cu-2 is occupied by one electron. 
A more complete description includes all five $d$ orbitals from each Cu ion, resulting in a ten-orbital model (blue solid line in Fig.~\ref{fig:dft-result}(c)). 
States derived from Y, Sb, and O lie at higher binding energies and do not significantly contribute to the low-energy effective model. 

The conclusion that the Cu-1 and Cu-2 $d$ orbitals are integer-filled holds only in the atomic limit. Upon forming Bloch wavefunctions—particularly after downfolding to effective low-energy models—the orbital filling often deviates substantially from integer values.
Nonetheless, our DFT calculations reveal that the electron occupation of the twofold degenerate $E_g$ bands at the Cu-1 site remains 3, whereas the $A_{1g}$ state at the Cu-2 site is occupied by approximately 1.15 electrons. This deviation likely arises from stronger Cu–O hybridization associated with an additional shorter Cu–O bond (see Fig.~\ref{fig:dft-result}(b)).

\begin{figure*}[t]
	\centering
	\includegraphics[width=\linewidth]{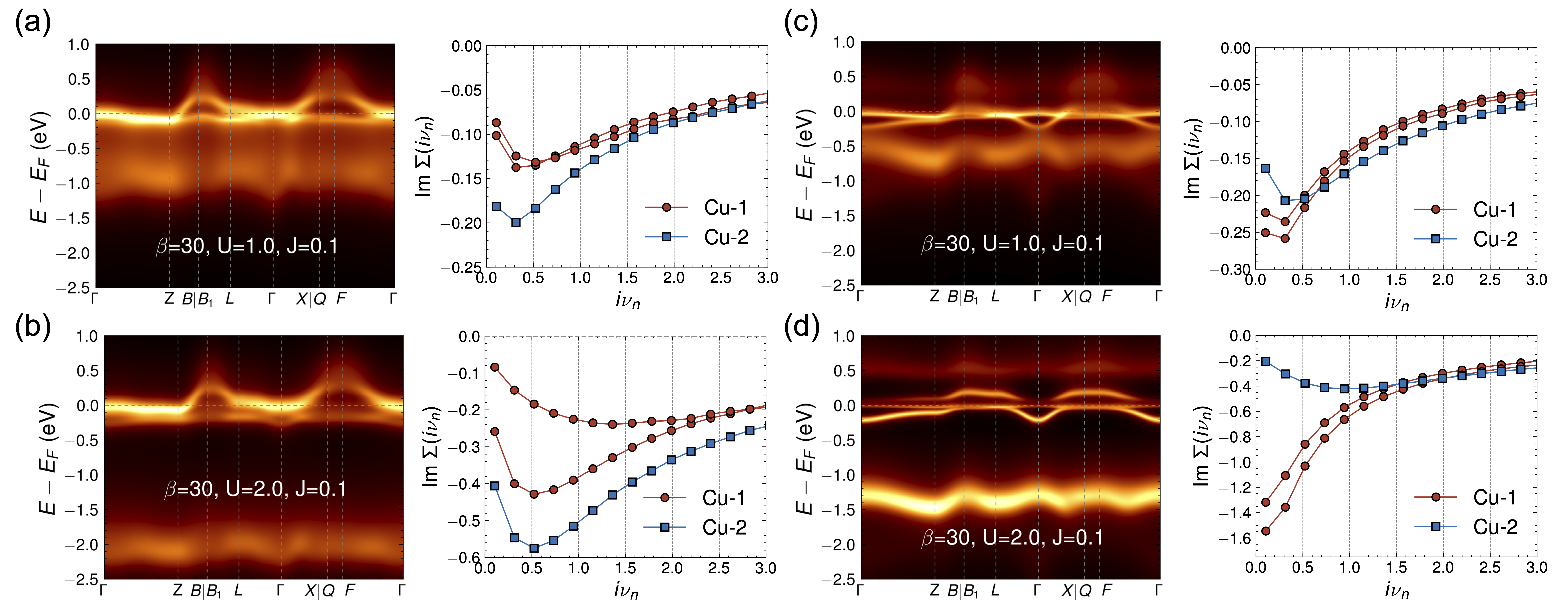}
	\caption{\textbf{DMFT spectral function and self-energy.}  (a, b) and (c, d) correspond to two different types of DMFT impurity construction. In (a, b), all three orbitals depicted in Fig.~\ref{fig:dft-result}(d) are treated within one impurity, while in (c, d) Cu-1 and Cu-2 are considered to be independent and each forms a DMFT impurity problem. In all four calculations, $\beta$ = 30, $J$ = 0.1 eV. $U$ are set as 1 eV in (a, c), and 2 eV in (b, d). }
	\label{fig:dmft}
\end{figure*}

\subsection{Correlated Electronic Structure}
Given the distinct filling levels of the two Cu ions and the narrow bandwidth near the Fermi level, electronic correlations are expected to behave differently in shaping the states of the electrons around the two Cu ions. 
In this section, we investigate the evolution of the electronic structure under the influence of Coulomb interactions. To this end, we employ the DFT+DMFT scheme implemented in our in-house ``Package for Analyzing Correlated Systems'' (PACS)~\cite{PhysRevB.85.115103, PACS_1} and cross-validate our results using the TRIQS package~\cite{parcollet2015triqs, aichhorn2016triqs}. In both implementations, we restrict the correlated subspace to the three partially filled bands near the Fermi level (indicated by yellow solid lines in Fig.~\ref{fig:dft-result}(d)).

Since DMFT is a local approximation, careful consideration must be given to the construction of the impurity problem. The two Cu sites experience distinct local crystal fields. 
Including all three Wannier functions in a single impurity problem would mix the two independent Cu ions via electronic correlations, potentially obscuring their distinct physical characters. A more appropriate approach is to define two separate impurity problems, one for each Cu site. In the following, we examine both schemes and compare their results in Fig.~\ref{fig:dmft}.

In the first scheme, we treat all three Wannier functions within a single impurity problem. In addition to the intra-site couplings between the two degenerate orbitals on Cu-1, the $A_{1g}$ state on Cu-2 is also coupled to Cu-1 via single-particle hybridizations and density–density Coulomb interactions, resulting in an effective three-orbital impurity model. We solved this model for two interaction strengths, $U = 1.0$ and $2.0$ eV, with $\beta = 30$ and $J = 0.1$ eV. As shown in Fig.~\ref{fig:dmft}(a), the resulting correlated spectral function exhibits broad features spanning from $-1.5$ eV to $1.0$ eV. The imaginary parts of the self-energies for the two Cu sites show distinct behaviors. The two $E_g$ Wannier states on Cu-1 remain nearly degenerate (red line with circles) and exhibit a smaller self-energy than that of the $A_{1g}$ state on Cu-2 (blue line with squares), indicating weaker band renormalization for Cu-1. However, increasing $U$ to $2.0$ eV lifts the degeneracy of the $E_g$ states, resulting in two different self-energies for Cu-1. In this unified impurity treatment, electronic correlations enhance the single-particle hybridization among the three Wannier states, leading to an unreasonable breaking of the $E_g$ symmetry.

A more physically sound DMFT framework assigns the three Wannier states to separate impurity problems based on their spatial localization. In the second scheme, we therefore treat the two $E_g$ states on Cu-1 and the $A_{1g}$ state on Cu-2 in two independent impurity problems, solving for their respective self-energies separately. 
Both self-energies are incorporated simultaneously into the DMFT self-consistency loop and the DFT+DMFT upfolding/downfolding procedure. 
The resulting spectral functions and self-energies are presented in Fig.~\ref{fig:dmft}(c) and (d). First, similar to the unified treatment, the system remains metallic across the parameter range studied. 
However, the self-energy now preserves the expected symmetry: the $E_g$ states remain nearly degenerate and distinct from the $A_{1g}$ state at both interaction strengths. 
Second, we observe that the integer-filled $E_g$ states are more sensitive to electronic correlations. As $U$ increases from 1 to 2 eV, the self-energy of the $E_g$ states evolves from metallic toward more insulating behavior, while that of the $A_{1g}$ state retains its metallic character. This stark contrast arises from the difference in orbital fillings: the integer-filled $E_g$ states on Cu-1 experience stronger band renormalization, whereas the partially filled $A_{1g}$ state remains relatively weakly correlated. 
Consequently, Y$_3$Cu$_2$Sb$_3$O$_{14}$ exhibits a site-selective band renormalization, where bands derived from Cu-1 orbitals are more strongly correlated than those from Cu-2.
The observation of site-selective band renormalization in Y$_3$Cu$_2$Sb$_3$O$_{14}$ may offer a promising explanation to the two-step spin condensation observed in experiment, where the charge degree of Cu-2 is less correlated and only freeze at much lower temperature, inhibiting the condensation of spins around Cu-2. 

This site-selective behavior is reminiscent of the extensively studied rare-earth nickelates $R$NiO$_3$ (with $R =$ Sm, Eu, Y, Lu), in which Ni occupies two distinct sublattices characterized by different Ni–O bond lengths~\cite{PhysRevLett.109.156402}. In those materials, Ni ions in long-bond octahedra adopt a nominal $d^8$ configuration, whereas those in short-bond octahedra exhibit a $d^8\underline{L}^2$ configuration, where two ligand holes compensate two electrons in the Ni $e_g$ orbitals. This leads to inequivalent Ni sites and differential band renormalization under electronic correlations~\cite{PhysRevB.91.075128, PhysRevB.92.155145, PhysRevB.102.085155, PhysRevB.103.085110}. In Y$_3$Cu$_2$Sb$_3$O$_{14}$, the situation is even more striking: the two Cu$^{2+}$ ions experience completely inverted crystal fields, giving rise to distinct correlation responses in the two Cu sites.

\subsection{Spin Susceptibility}

\begin{figure*}[htbp]
	\centering
	\includegraphics[width=\linewidth]{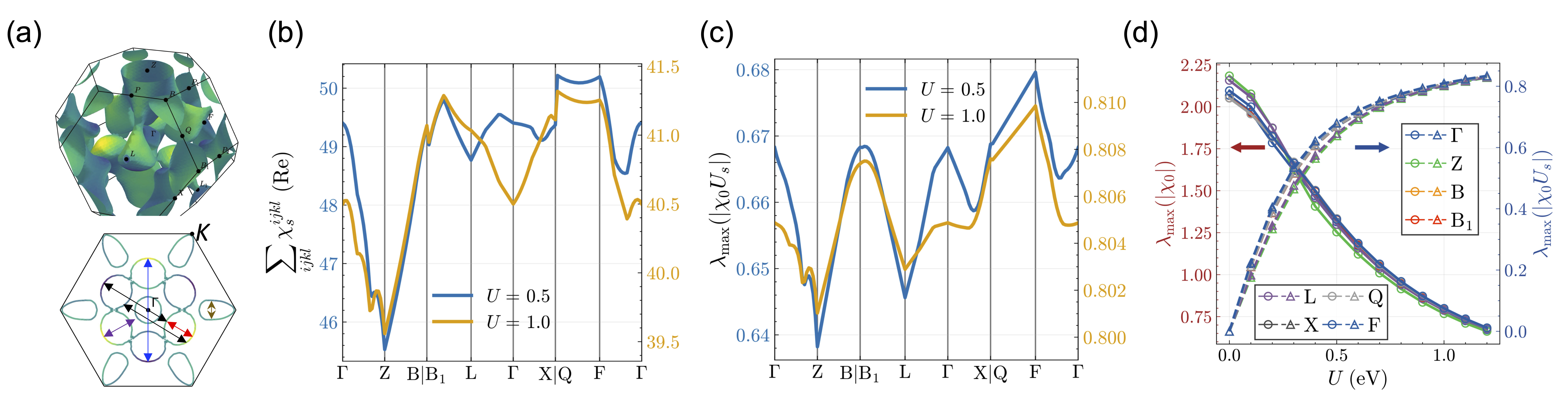}
	\caption{\textbf{Fermi surface and spin susceptibility $\chi(\bm{q}, i\omega_m=0)$.} 
	(a).  The Fermi surface in 3D BZ and in a 2D cut through $\Gamma$ point display multiple pockets and nesting structures. 
	(b). The spin susceptibility calculated from Eq.~\eqref{Eq:chis} displays multiple peaks with only small variation, which is further suppressed by electronic correlations. The blue and yellow curves correspond to different interaction parameters as shown by the legend. 
	(c). The leading eigenvalues of $\chi_0 U_s$ follow a similar structure as $\chi_s$ in (b), whose variation is also reduced by electronic correlations. 
	(d). Both the leading eigenvalues of $\chi_0$ (the left-yaxis in red) and $\chi_0 U_s$ (the right-yaxis in blue) for different momentum wave vectors fall into a single curve under the evolution of electronic correlations, with the former decreases and the latter increases with $U$.   
	}
	\label{fig:chi+nesting}
\end{figure*}
To further shed light on the absence of long-range magnetic order in  Y$_3$Cu$_2$Sb$_3$O$_{14}$, in this section we examine the spin susceptibility and its evolution under the influence of Coulomb interactions. The spin susceptibility, which encodes the magnetic response of the system, often serves as a key indicator of spontaneous symmetry breaking. 
In particular, a divergent spin susceptibility at a specific wave vector directly reveals the dominant real-space spin correlation pattern associated with that wave vector. Given the absence of long-range magnetic order down to very low temperatures in Y$_3$Cu$_2$Sb$_3$O$_{14}$, we anticipate a spin susceptibility that remains relatively broad and structureless across the entire BZ.

We compute the spin susceptibility using the ten-band effective model shown in Fig.~\ref{fig:dft-result}(c), which provides a better description of the DFT Fermi surface and the localized nature of the $d$ orbitals.
The spin susceptibility is evaluated within two distinct schemes. In the first approach, we calculate the susceptibility from the convolution of two single-particle Green's functions as
\begin{eqnarray}\label{Eq:bubble}
	\chi^{0}_{il;jk}(\bm{q}, i\omega_m) = && -\frac{1}{\beta N_{\bm{k}}} \nonumber \\ \sum_{\bm{k},i\nu_n}G_{ij} (\bm{k},&&i\nu_n)G_{kl}(\bm{k}+\bm{q},i\nu_n+i\omega_m)\;,
\end{eqnarray}
where $i,j,k,l$ denote orbital indices. Here, $G_{ij}(\bm{k}, i\nu_n)$ is the dressed single-particle Green's function obtained from a converged FLEX calculation. Thus, $\chi^0_{il;jk}(\bm{q}, i\omega_m)$ differs from the non-interacting susceptibility; while both share the same convolution structure, the bubble susceptibility in Eq.~\eqref{Eq:bubble} incorporates renormalization effects from the single-particle self-energy. In the second scheme, we further renormalize the spin susceptibility within the random phase approximation in the magnetic channel:
\begin{eqnarray}\label{Eq:chis}
	\chi_s = \frac{\chi_0}{1-\chi_0 U_s}\;.
\end{eqnarray}

From the Lehmann representation of Eq.~\eqref{Eq:bubble}, it is evident that the Fermi surface topology plays a crucial role in determining the divergent behavior of the spin susceptibility. 
When a substantial number of electrons on the Fermi surface scatter with the same wave vector, the resulting coherence can lead to a divergence of the spin susceptibility at that wave vector. 
Figure~\ref{fig:chi+nesting}(a) displays the three-dimensional Fermi surface, with a two-dimensional cut through the $\Gamma$ point shown on the right. The Fermi surface of Y$_3$Cu$_2$Sb$_3$O$_{14}$ consists of several pockets of comparable size, giving rise to multiple competing nesting vectors. A few representative nesting vectors are indicated in the two-dimensional plot, each potentially contributing comparable coherence to electron scattering. As a result, the spin susceptibility exhibits a multi-peak structure with several competing maxima.

Figure~\ref{fig:chi+nesting}(b) presents the spin susceptibility calculated from Eq.~\eqref{Eq:chis} along a selected momentum path, comparing results for two different values of $U$. The blue and yellow curves correspond to $U = 0.5$ eV and $U = 1.0$ eV, respectively. For $U = 0.5$ eV, the total spin susceptibility summed over all orbital combinations, $\chi_s^\text{tot}$, exhibits several peaks, yet its maximum variation remains only about 10\% of its absolute value. Hence, $\chi_s^\text{tot}$ retains a broad, nearly featureless profile, indicating the absence of a dominant divergent wave vector in this system. Upon increasing $U$ to 1.0 eV, the overall magnitude of $\chi_s^\text{tot}$ is suppressed (see the right axis in yellow), and the relative variation further decreases to approximately 5\%. This trend suggests that electronic correlations enhance the featureless character of the spin susceptibility.

The growing competition among different wave vectors in $\chi_s(\bm{Q})$ is further confirmed in the leading eigenvalues of $U_s\chi_0(\bm{Q})$, which appear in the denominator of Eq.~\eqref{Eq:chis} and drive the divergence of $\chi_s(\bm{Q})$ as they approach unity. Figure~\ref{fig:chi+nesting}(c) shows these leading eigenvalues for the same interaction parameters used in panel (b). The variation of $\lambda_{\text{max}}(U_s\chi_0(\bm{Q}))$ across different $\bm{Q}$ is about 6\% for $U = 0.5$ eV, but reduces to 1.25\% when $U$ is increased to 1.0 eV, despite the absolute value rising from 0.68 to 0.82. Consistent with the behavior of $\chi_s(\bm{Q})$, increasing interaction strength suppresses the momentum dependence of $\lambda_{\text{max}}(U_s\chi_0(\bm{Q}))$, leading to a nearly simultaneous divergence tendency at all $\bm{Q}$.

Figure~\ref{fig:chi+nesting}(d) extends this analysis to more values of $U$. 
In addition to $U_s\chi_0(\bm{Q})$ (right y-axis, blue), we also show the leading eigenvalues of $\chi_0(\bm{Q})$ (see Eq.~\eqref{Eq:bubble} and the left y-axis of Fig.~\ref{fig:chi+nesting}(d) in red), which appear in the numerator of Eq.~\eqref{Eq:chis}. 
As $U$ increases, the leading eigenvalues of $U_s\chi_0(\bm{Q})$ at characteristic wave vectors grow with nearly identical magnitudes. Concurrently, those of the bubble susceptibility $\chi_0(\bm{Q})$ decrease. Notably, for both quantities, the leading eigenvalues across all $\bm{Q}$ considered collapse onto nearly the same curve, indicating that no single magnetic wave vector dominates. The system thus exhibits strong competition among various magnetic configurations, precluding the establishment of long-range magnetic order with a well-defined wave vector $\bm{Q}$.

\section{Discussion and conclusions}\label{sec:discussions-and-conclusions}

The theoretical results presented in this work support Y$_3$Cu$_2$Sb$_3$O$_{14}$ as a compelling candidate for hosting a paramagnetic ground state. 
The combination of geometric frustration inherent to the triangular lattice geometry, the presence of two inequivalent Cu sites with inverted crystal-field splittings, and the strong electronic correlations revealed by our DMFT calculations creates an ideal environment for the suppression of long-range magnetic order.

Our DFT analysis demonstrates that the two distinct Cu sites exhibit fundamentally different orbital configurations. 
Cu-1, with its trigonal distorted coordination, follows the conventional crystal-field hierarchy with the $d_{z^2}$ orbital lowest in energy. In contrast, Cu-2 experiences an extreme axial compression that inverts this ordering, placing the unpaired hole in the $d_{z^2}$ orbital. This inversion, combined with the three-dimensional network of triangular layers, generates a complex electronic structure characterized by narrow bands with different filling levels.

The DMFT calculations reveal a site-selective band renormalization at Cu-1 and Cu-2, which is deeply related to the distinct local environments of the two Cu ions. When the two ions are correctly treated in separate DMFT impurity problems, the symmetry of the twofold $E_g$ states at Cu-1 is preserved; these states are close to a Mott transition, while the $A_{1g}$ state at Cu-2 remains metallic across the parameter range studied.
The contrasting electronic responses of the low-energy states at Cu-1 and Cu-2 may provide an explanation for the two-step spin condensation observed experimentally. The more correlated charge degrees of freedom at Cu-1 localize at a relatively higher temperature, allowing the remaining spin degrees of freedom on Cu-1 to interact and condense. In contrast, the charge at Cu-2 remains highly fluctuating until very low temperatures preventing the spins from condensing. 
The DMFT local approximation cannot fully open the charge gap. 
If the system is insulating from experimental point of view, a short-range magnetic correlation may play a critical role in this system. 

Another significant observation is that our spin susceptibility calculations within the FLEX approximation provide direct evidence for the suppression of magnetic order. The absence of a dominant peak in $\chi_s(\bm{Q})$, together with the nearly $\bm{Q}$-independent behavior of the leading eigenvalues of $U_s\chi_0(\bm{Q})$, indicates that multiple magnetic instabilities compete with nearly equal strength. 
This competition becomes more pronounced with increasing $U$ and the variation of $\lambda_{\text{max}}$ across different wave vectors is further suppressed. 
Such behavior is a hallmark of a highly frustrated magnet where magnetic order is unable to select a unique ground state.

These findings have important implications for experimental efforts to identify and characterize QSL behavior in this material. 
The two-step spin condensation observed experimentally \cite{saha2025} finds a natural explanation in our theoretical framework: the hierarchy of energy scales introduced by the two distinct Cu sites and their competing interactions may give rise to a cascade of fractionalization as temperature is lowered. 
Future experiments should focus on detecting the characteristic signatures of the electronic and magnetic correlations related to the two different Cu local environment.

\begin{acknowledgements}
This work was supported by the National Key R\&D Program of China (No. 2022YFA1402703), the National Nature Science Foundation of China (Grant Nos. 12574271, 92365204), Sino-German Mobility program (No. M-0006), and Shanghai 2021- Fundamental Research Area (No. 21JC1404700). Part of the calculations was performed at the HPC Platform of ShanghaiTech University Library and Information Services, and the School of Physical Science and Technology.
\end{acknowledgements}

\appendix
\section{Details of the Methods}
\subsection{DFT and DMFT calculations}
The first-principle calculations of Y$_3$Cu$_2$Sb$_3$O$_{14}$ were carried out in the framework of density-functional theory (DFT). 
We adopted the Quantum ESPRESSO (QE) implementation of DFT for the band structure calculations~\cite{giannozzi2009quantum, giannozzi2017advanced}.
The kinetic energy cutoff for wavefuncton and charge density was set as 80 Ry and 500 Ry, respectively. For the Brillouin-zone integrations, a Monkhorst–Pack~\cite{monkhorst1976special} $k$-point mesh of $9\times9\times9$ was used. 
The sg15 database~\cite{PhysRevB.88.085117} of optimized normconserving Vanderbilt PPs (ONCV) was adopted for the non-relativistic pseudopotential. 

To investigate the correlated electronic structure, we performed one-shot DFT+DMFT calculations by downfolding the DFT electronic structure to a three-band effective model using the maximally localized Wannier Functions(WFs)~\cite{marzari1997maximally, souza2001maximally} with the pw2wannier interface~\cite{pizzi2020wannier90}.  
The DMFT impurity problem with density-density interactions was solved using the continuous-time hybridization expansion solver~\cite{werner_continuous-time_2006, werner_hybridization_2006} implemented in home-made code ``Package for Analyzing Correlated System " (PACS)~\cite{PhysRevB.85.115103, PACS_1} and further cross-checked with TRIQS packages~\cite{parcollet2015triqs, aichhorn2016triqs, seth2016triqs}. 
The self-energy in real frequency was obtained by analytic continuation based on the maximum entropy~\cite{gubernatis1991quantum, triqsmaxent}.

\subsection{Fluctuation exchange approximation}

The spin susceptibility was evaluated with multi-orbital Fluctuation Exchange (FLEX) approximation, introduced by Bickers et al. \cite{bickers2004self}. 
It is a perturbative diagrammatic approach that self-consistently treats spin, charge, singlet, and triplet fluctuations. 
In this work, we considered the ten-band tight-binding model obtained from Wannier90 with density-density multi-orbital interaction:
\begin{equation}\label{equ:int}
	\begin{aligned}
		H &= H_0 + H_{\text{int}} \\
		H_0 &= \sum_{i,j,\alpha,\beta,\sigma}(-t_{ij}^{\alpha\beta}-\mu\delta_{ij}\delta^{\alpha\beta})c^{\dagger}_{i\alpha\sigma}c_{j\beta\sigma} \\
		H_{\text{int}}  &= \frac{1}{4} \sum_{i,1234} U_{\sigma_1\sigma_2\sigma_3\sigma_4}^{a_1a_2a_3a_4}c^{\dagger}_{ia_1\sigma_1}c_{ia_2\sigma_2}c^{\dagger}_{ia_4\sigma_4}c_{ia_3\sigma_3} \\
	\end{aligned}
\end{equation}
$H_0$ is the ten-band (projected by Cu-$d$ orbitals) tight-binding model. $H_{\text{int}}$ is the simplified interaction Hamiltonian, expressed as:
\begin{eqnarray}
	H_{\text{int}} = U\sum_{a}\hat{n}_{a\uparrow}\hat{n}_{a\downarrow} + U^{\prime}\sum_{a\ne a^{\prime}}\hat{n}_{a\uparrow}\hat{n}_{a'\downarrow} \nonumber\\
	+ (U^{\prime}-J)\sum_{a<a^{\prime},\sigma}\hat{n}_{a\sigma}\hat{n}_{a^{\prime}\sigma}
\end{eqnarray}
By utilizing the SU(2) identity
\begin{eqnarray}
	\bm{\tau}_{\sigma_1\sigma_2}\cdot\bm{\tau}_{\sigma_4\sigma_3} = 2\delta_{\sigma_1\sigma_3}\delta_{\sigma_4\sigma_2} - \delta_{\sigma_1\sigma_2}\delta_{\sigma_3\sigma_4}
\end{eqnarray}
the interaction term in Eq.~\eqref{equ:int} can be expressed as
\begin{eqnarray}
	U_{\sigma_1\sigma_2\sigma_3\sigma_4}^{a_1a_2a_3a_4} = -U_m^{a_1a_2, a_4a_3}\delta_{\sigma_1\sigma_3}\delta_{\sigma_4\sigma_2}\nonumber\\
	+ \frac{1}{2}(U_d^{a_1a_2,a_4a_3}+U_m^{a_1a_2, a_4a_3})\delta_{\sigma_1\sigma_2}\delta_{\sigma_3\sigma_4}
\end{eqnarray}
After symmetrizing the interaction tensor in magnetic ($m$) and density ($d$) channels,
\begin{eqnarray}
	U_m = U_{\uparrow\uparrow\downarrow\downarrow} - U_{\uparrow\uparrow\uparrow\uparrow}, U_d = U_{\uparrow\uparrow\uparrow\uparrow} + U_{\uparrow\uparrow\downarrow\downarrow}\;,
\end{eqnarray}
one has
\begin{eqnarray}
	U_m=\left\{ 
	\begin{aligned}
		U &,\\
		U'-J&,\\
		J&,\\
		0&,\\
	\end{aligned}
	\right.
	U_d=\left\{ 
	\begin{aligned}
		U \quad, & \ 1=2=3=4\\
		J-U', & \ 1=3\neq2=4 \\
		2U'-J ,&\ 1=2\neq4=3 \\
		0\quad ,&\ 1=4\neq3=2\\
	\end{aligned}
	\right.
\end{eqnarray}
To simplify the notation, here we define the collective variable $1 \equiv (\bm{r}_1, \tau_1, a_1, \sigma_1)$.

Similarly, the interaction tensor can also be written in the singlet (s) and triplet (t) particle-particle channels. 
The parquet equation self-consistently connects these four channels to satisfy the crossing symmetry of the various two-particle vertices. 
With Schwinger-Dyson equation, the self-energy is constructed from the two-particle vertices as:
\begin{eqnarray}\label{equ:sigma}
	\Sigma_{2}(11') =&& \frac{1}{2} \{ \nonumber\\ 
	-\frac{1}{2} G(76)&&[\frac{1}{2}\Lambda^{\text{irr}}_\text{d} G^{\text{ph}} U_\text{d} +\frac{3}{2} \Lambda^{\text{irr}}_\text{m} G^{\text{ph}} U_\text{m}](17;1'6) \nonumber\\
	+  G(67)&&[\frac{1}{2}\Lambda^{\text{irr}}_\text{s} G^{\text{pp}} U_\text{s} + \frac{3}{2}\Lambda_\text{t}^{\text{irr}} G^{\text{pp}} U_\text{t}](17;1'6)\} \\
	- G(76) && \left[\frac{1}{2}\Phi_{\text{d}} G^{\text{ph}} U_{\text{d}} + \frac{3}{2}\Phi_{\text{m}} G^{\text{ph}} U_{\text{m}}\right](17;1'6) \nonumber\\
	+ G(67) && \left[\frac{1}{2}\Psi_{\text{s}} G^{\text{pp}} U_{\text{s}} + \frac{3}{2}\Psi_{\text{t}} G^{\text{pp}} U_{\text{t}}\right](17;1'6)\;, \nonumber
\end{eqnarray}
where $\Lambda$ and $\Phi/\Psi$ denote the two-particle fully irreducible and channel-dependent reducible contributions, respectively. 
\begin{eqnarray}\label{equ:phi}
	\Phi_r(12;34)=\sum_{i\omega_m}  &&e^{i\omega_m\tau} \\ \nonumber
	&& [\Lambda_r(\bm{1}-G^{ph}\Lambda_r)G^{ph}\Lambda_r](i\omega_m)
\end{eqnarray}
where $r=\text{d/m}$, with analogous expressions for $s$ and $t$ channels. 
FLEX approximate the fully irreducible vertex $\Lambda$ with the bare interaction $U$. 
We self-consistently solve Eqs.~\eqref{equ:sigma}-\eqref{equ:phi} with
\begin{eqnarray}\label{equ:dyson}
	G^{-1}(\bm{k},i\nu_n) = G_0^{-1}(\bm{k},i\nu_n) - \Sigma(\bm{k},i\nu_n) 
\end{eqnarray}
and the non-interacting Green's function
\begin{eqnarray}\label{equ:bareg}
	G_0(\bm{k},i\nu_n)=\frac{\bm{1}}{(i\nu_n+\mu)\bm{1}-H_{0}(\bm{k})} \;,
\end{eqnarray}
where the chemical potential $\mu$ is adjusted in each self-consistent iteration to maintain a fixed electron density $n$. 

In our implementation, we adopt the \texttt{sparse-ir} basis\cite{shinaoka2017, shinaoka2020sparse} to compress the representation of Green's functions. 
Computational efficiency is further enhanced by using \texttt{mpi4py-fft}\cite{mpi4py-fft}. 
With all these considerations, we are able to solve the ten-band model in Fig.~\ref{fig:chi+nesting} with $\beta$=30, $U$=2 eV, and a $6\times6\times6$ $\bm{k}$-mesh within 100 CPU hours.

\bibliographystyle{apsrev4-2}
\bibliography{ref}

\end{document}